\documentclass[12pt]{article}
\pdfoutput=1
\usepackage{putex}
\usepackage{feyn}
\usepackage[vcentermath]{youngtab}
\usepackage{subfig}
\usepackage{lscape}

\usepackage{graphicx}
\usepackage{epstopdf}
\usepackage{enumerate}
\usepackage{cite}
\usepackage{tensor}
\usepackage{slashed}
\usepackage{amsmath}
\usepackage{amssymb}
\usepackage{mathrsfs}
\usepackage{lgrind}

\usepackage{bbm}

\usepackage{hyperref}

\numberwithin{equation}{section}

\newcommand {\be} {\begin {equation}}
\newcommand {\ee} {\end {equation}}

\newcommand {\bes} {\begin {equation*}}
\newcommand {\ees} {\end {equation*}}

\newcommand{\eps}{\epsilon}

\def\CH{{\cal H}}

\newcommand{\beq}{\begin{equation}}
\newcommand{\eeq}{\end{equation}}

\def\be{ \begin{equation} }
\def\ee{ \end{equation} }

\def\Tr{{\textrm{Tr}}}
\begin{document}

\preprint{PUPT-2508}

\institution{PU}{Department of Physics, 
 Princeton University, Princeton,  NJ 08544 }

\title{
On  Large $N$ Expansion of \\ the  Sphere Free Energy 
}


\authors{Grigory Tarnopolsky\worksat{\PU}}

\abstract{ We propose formulas for the $1/N$ correction to  the sphere free energy  of  theories with 4-fermion interactions, which are conformal for $d>2$. We also propose a formula for the scalar $O(N)$ model.  Expanding these formulas  in small $\eps$ near various integer dimensions we find a  perfect agreement with results obtained using  $\eps$-expansion technique. In $d=3$, the large $N$ results with the $1/N$ correction included are in  good agreement with  the Pade resummed $\eps$-expansion.
}

\date{}
\maketitle

\tableofcontents

\section{Introduction and Summary}

The  free energy of a Quantum Field Theory (QFT) on a $d$-dimensional sphere $F=-\log Z_{S^d}$  is an extensive quantity, therefore the leading term of it is proportional to the volume of the sphere. The value of this leading term is divergent and depends on the regularization scheme. For 3-d Conformal Field Theory (CFT)  the non-divergent constant part of the sphere free energy is universal and does not depend on the sphere radius $R$.  We call this part  simply the sphere free energy $F$. Therefore the sphere free energy $F$ of a CFT is a number. This number ''feels'' the whole space and is non-local in its nature. The interest in $F$ is due to the $F$-theorem \cite{Jafferis:2011zi} , which suggests that $F$  is a measure of the degrees of freedom of a CFT. 
More precisely the $F$-theorem implies that an RG flow from a CFT$_{1}$ to a  CFT$_{2}$
 is only possible  if 
 \begin{align}
 F_{\textrm{CFT}_{1}}>F_{\textrm{CFT}_{2}}\,.
 \end{align}
Knowledge of $F$  helps to rule out some possible RG flows and support  others. 
The other interesting aspect of the sphere free energy is that it can be  formulated in terms of the
entanglement entropy across a circle \cite{Myers:2010xs,Casini:2011kv}.
The three-dimensional $F$-theorem has been proved  using properties of the entanglement entropy in relativistic theories \cite{Casini:2012ei} (see also \cite{Liu:2012ee}).

Unfortunately in the majority of  strongly coupled 
theories it is not possible to compute $F$ exactly. Only in some specific supersymmetric models the sphere free energy  can be computed explicitly using localization technique \cite{Kapustin:2009kz, Jafferis:2010un,Giombi:2014xxa}.
Thus it is important to estimate $F$ in different strongly coupled conformal theories. The sphere free energy for the $O(N)$ scalar, Gross-Neveu, Nambu-Jona-Lasinio and QED models  was estimated in series of works \cite{Giombi:2014xxa,Fei:2015oha,Fei:2016sgs, Giombi:2015haa}  using the Wilson-Fisher $\eps$-expansion. This computation is subtle and involves a careful treatment of  curvature terms, which  contribute to  the sphere free energy at the critical point. The advantage of the  $\eps$-expansion  is that it works quite well for an arbitrary $N$ and gives a rather precise estimate for $F$ at $d=3$.

Another standard approximation technique is the large $N$ expansion \cite{Wilson:1972zzb,Brezin:1972se}. It was applied to the sphere free energy $F$ in several papers \cite{Klebanov:2011gs, Klebanov:2011td,Giombi:2015haa, Diaz:2007an} but only up to the $N^{0}$ order. At this order one has to compute  the determinant of a special 
operator on a sphere.  

In the present paper we  compute the next order in $1/N$ in four different models, namely in the  Gross-Neveu, Nambu-Jona-Lasinio model with $U(1)$ chiral symmetry, Nambu-Jona-Lasinio model with $SU(2)$ chiral symmetry  and Wilson-Fisher $O(N)$ model.  All these models are CFTs in $2<d<4$. 

The Gross-Neveu (GN) model \cite{Gross:1974jv} in the flat space  has the action
\begin{align}
S_{\textrm{GN}}= \int d^{d}x  \Big( \bar{\psi}_{i}  {\not\,}\partial \psi^{i}+\frac{g}{2} (\bar{\psi}_{i}\psi^{i})^{2} \Big)  \,,
\label{GNorig}
\end{align}
where $\psi^{i}$, $i=1,\dots, N_{f}$ are the Dirac fermions. 
This model possesses the discrete chiral symmetry $\psi_i \rightarrow \gamma_5 \psi_i$. 
In $d=2$ this theory is asymptotically free for $N>2$  \cite{Gross:1974jv} and it has a unitary UV fixed point in $2<d<4$,  which describes the second-order phase transition where the
discrete chiral symmetry is broken.  

The Nambu-Jona-Lasinio models with $U(1)$ and $SU(2)$ chiral symmetries  \cite{Nambu:1961tp}  are described by the actions
\begin{align}
&S^{U(1)}_{\textrm{NJL}}=  \int d^{d}x  \Big( \bar{\psi}_{i}  {\not\,}\partial \psi^{i} +\frac{ g}{2} \left ( (\bar{\psi}_{i}\psi^{i})^{2}- (\bar{\psi}_{i}\gamma_5\psi^{i})^{2}\right )\Big) \,
,  \label{NJLU1orig}\\
&S^{SU(2)}_{\textrm{NJL}}=  \int d^{d}x  \Big( \bar{\psi}_{i\alpha}  {\not\,}\partial \psi^{i\alpha} +\frac{ g}{2} \left ( (\bar{\psi}_{i\alpha}\psi^{i\alpha})^{2}- (\bar{\psi}_{i\alpha}\gamma_5 \vec{\tau}_{\alpha\beta}\psi^{i\beta})^{2}\right )\Big) \,
,
\label{NJLSUorig}
\end{align}
where $\psi^{i}$ and $\psi^{i\alpha}$, $i=1,\dots, N_{f}$, $\alpha=1,2$ are the 4-component Dirac fermions, and $\tau^{a}$, $a=1,2,3$ are the Pauli matrices. For  $\gamma^{5}$  we use conventions $\{\gamma^{\mu},\gamma^{5}\}=0$ and $(\gamma^{5})^{2}=\bold{1}$.

In $d>2$ the four-fermion interactions (\ref{GNorig}) and (\ref{NJLU1orig}), (\ref{NJLSUorig}) are non-renormalizable, while they are renormalizable in the sense of the $1/N$ expansion \cite{Gross:1975vu, Parisi:1975im, Rosenstein:1988pt, Gat:1990xi}.
At finite $N$  the UV completion of these theories for $d<4$ are given by the Gross-Neveu-Yukawa,  Nambu-Jona-Lasinio-Yukawa and Gell-Mann-Levy models, considered originally in \cite{Hasenfratz:1991it,ZinnJustin:1991yn, GellMann:1960np}.

Using a Hubbard-Stratonovich transformation and  introducing  auxiliary fields we find that the critical point of the models  (\ref{GNorig}),  and   (\ref{NJLU1orig}), (\ref{NJLSUorig}) on the sphere in the large $N$ approach is described by the actions 
\begin{align}
S_{\textrm{GN}}&= \int d^{d}x \sqrt{g} \Big(\bar{\psi}_{i}\slashed{\nabla}\psi^{i}+\frac{1}{\sqrt{N}}\sigma \bar{\psi}_{i} \psi^{i} \Big)\,, \notag\\
S^{U(1)}_{\textrm{NJL}}&= \int d^{d}x \sqrt{g} \Big(\bar{\psi}_{i}\slashed{\nabla}\psi^{i}+\frac{1}{\sqrt{N}}\bar{\psi}_{i}(\phi_{1}+i\phi_{2}\gamma^{5})\psi^{i} \Big)\,, \notag\\
S^{SU(2)}_{\textrm{NJL} }&= \int d^{d}x \sqrt{g} \Big(\bar{\psi}_{i\alpha}\slashed{\nabla}\psi^{i\alpha}+\frac{1}{\sqrt{2N}}\bar{\psi}_{i\alpha}(\pi_{0}\delta_{\alpha\beta}+i\tau^{a}_{\alpha\beta}\pi_{a}\gamma^{5})\psi^{i\beta} \Big)\,, 
\end{align}
where $\sigma$, $\phi_{1}$, $\phi_{2}$ and $\pi_{0},\dots, \pi_{3}$ are  the scalar auxiliary fields and  $N = N_{f}\Tr \bold{1}$,  where  $\Tr \bold{1} = 4$ is the trace of the identity matrix in the spinor space.

\begin{figure}[h!]
    \centering
    \subfloat{{\includegraphics[width=7.5cm]{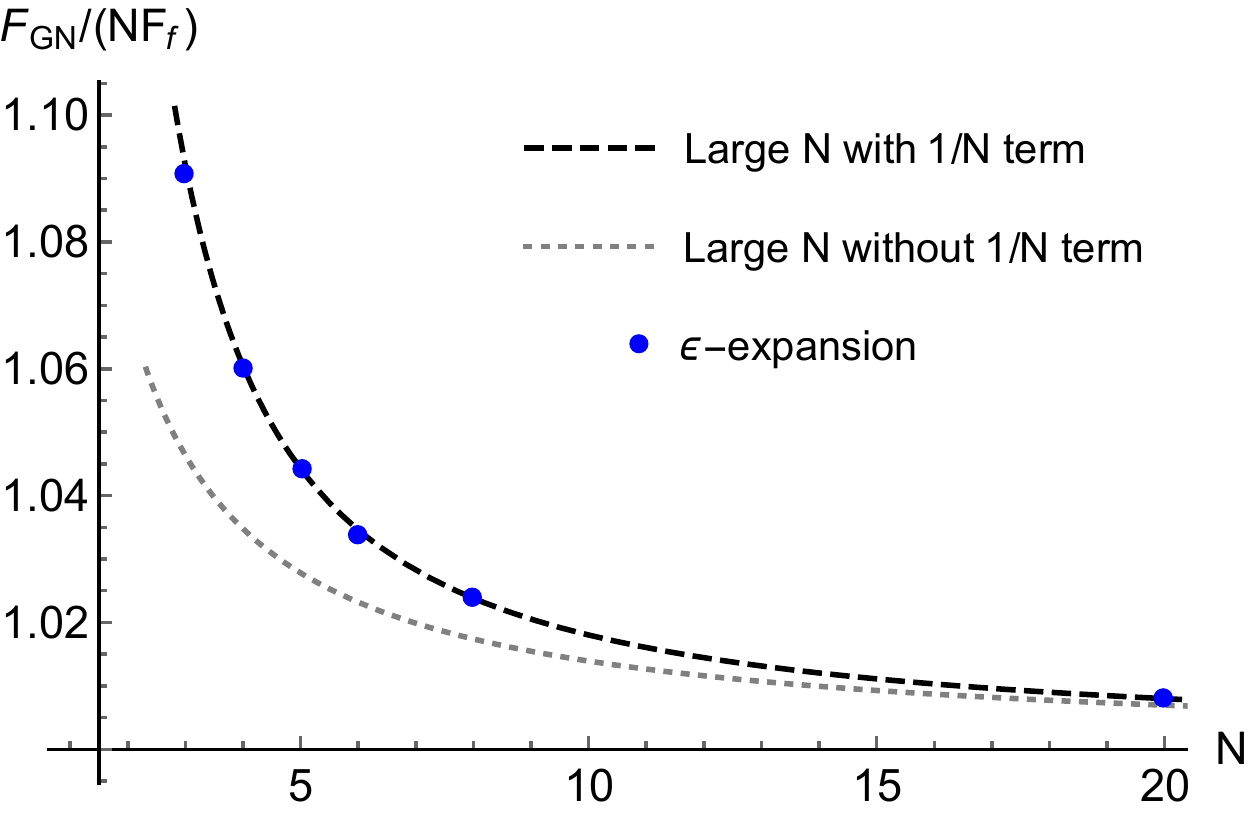} }}%
    \qquad
    \subfloat{{\includegraphics[width=7.5cm]{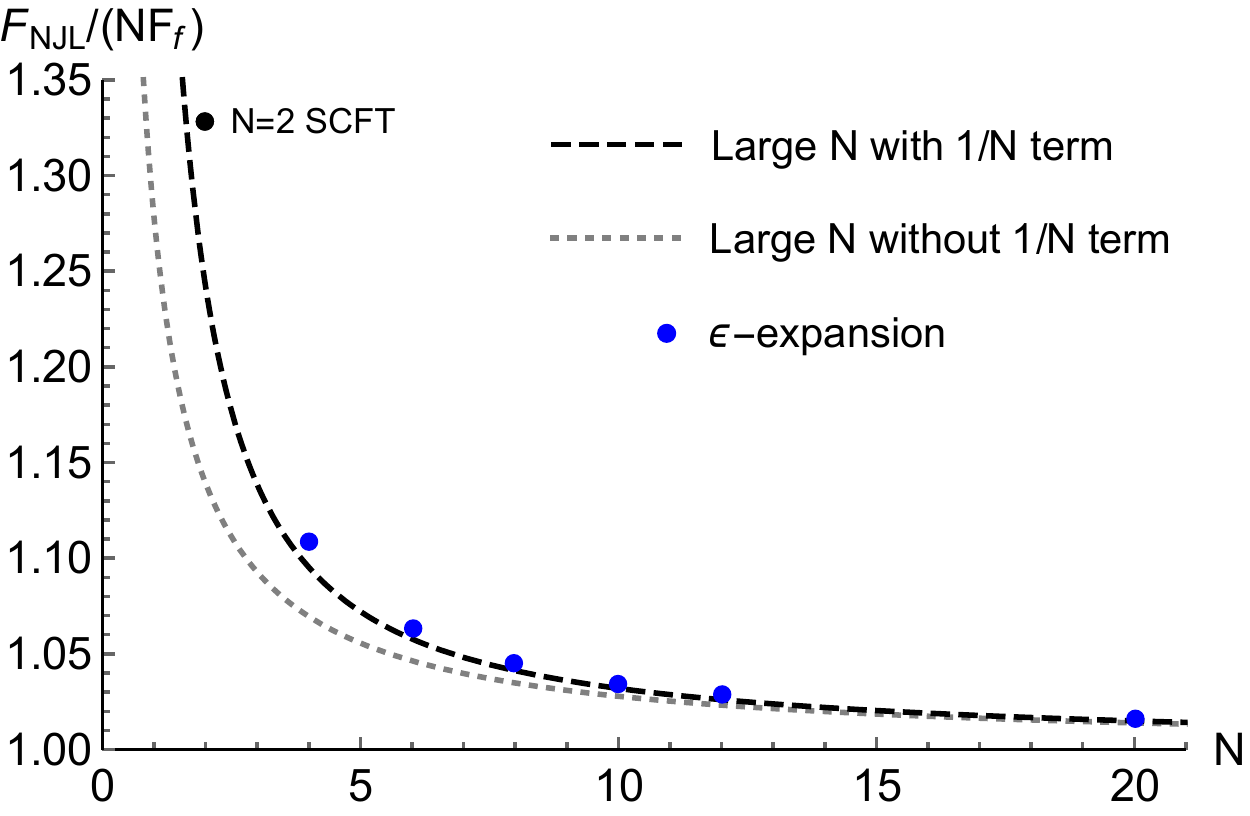} }}%
    \caption{The large $N$ and $\eps$-expansion for $F/(N F_{f})$  of the GN and NJL $U(1)$ models in $d=3$. The dashed black line depicts the large $N$ curve including the $1/N$ term, whereas the dotted gray line represents the large $N$ result without the $1/N$ correction. Blue dots stand for the $\eps$-expansion  obtained in \cite{Fei:2016sgs}. The black dot is the exact result for the Wess-Zumino model at $d=3$ obtained via localization technique in \cite{Jafferis:2010un}:  $F_{\textrm{WZ}}/(2F_{f})\approx 1.32806$. }
    \label{OneNvsEps}
\end{figure}

\begin{figure}[h!]
                \centering
                \includegraphics[width=8cm]{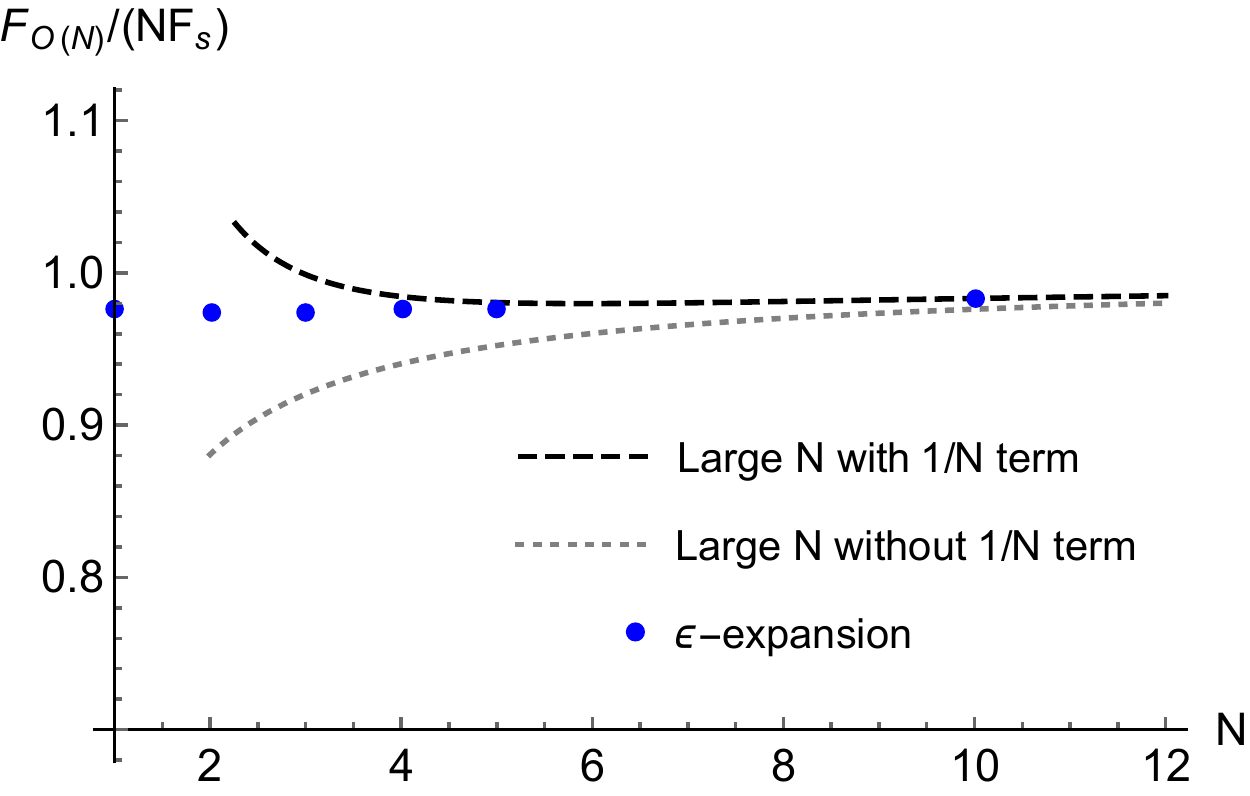}
                 \caption{The large $N$ and $\eps$-expansion for $F/(N F_{s})$ of the $O(N)$ scalar model in $d=3$. The dashed black line depicts the large $N$ curve including the $1/N$ term, whereas the dotted gray line represents the large $N$ result without the $1/N$ correction. Blue dots stand for the $\eps$-expansion  obtained in \cite{Fei:2015oha}.}
\label{OneNvsEpsON}
\end{figure}

The critical point of  the Wilson-Fisher $O(N)$ model  at the large $N$ is described by the action 
\begin{align}
S_{O(N) } &= \int d^{d}x \sqrt{g} \Big(\frac{1}{2}(\partial_{\mu} \phi^{i})^{2}+\frac{d-2}{8(d-1)} \mathcal{R}\phi^{i}\phi^{i}+\frac{1}{2\sqrt{N}}\sigma \phi^{i}\phi^{i}\Big)\,,
\end{align}
where $\phi^{i}$, $i=1,\dots, N$ are scalar fields, $\sigma$ is the scalar auxiliary field and $\mathcal{R} =d(d-1)/R^{2}$ is the scalar curvature of the sphere.  Our main proposals  are 
\begin{align}
F_{\textrm{GN}} &= NF_{f} + \delta F_{\Delta=d-1}
+ \frac{1}{N} \frac{1}{d} \gamma_{\psi,1}^{\textrm{GN}}+\mathcal{O}(1/N^{2})\,,  \notag\\
F_{\textrm{NJL}}^{U(1)} &= NF_{f} +  2\delta F_{\Delta=d-1}
+ \frac{1}{N} \frac{d-2}{d(d-1)} \gamma_{\psi,1}^{\textrm{NJL\,} U(1)}+\mathcal{O}(1/N^{2})\,,  \label{Fres1}\\
F_{\textrm{NJL}}^{SU(2)} &= 2 NF_{f} +  4\delta F_{\Delta=d-1}
 +\frac{1}{N} \frac{d-4}{d(d-1)} \gamma_{\psi,1}^{\textrm{NJL\,} SU(2)}+\mathcal{O}(1/N^{2})\, \notag
 \end{align}
and   
 \begin{align}
 F_{O(N)} &= NF_{s} + \delta F_{\Delta=d-2}
+ \frac{1}{N} \Big(\frac{1}{d}+\frac{1}{3 (d-2)}-\frac{2}{3 (d-1)}\Big)\gamma_{\phi, 1}^{\textrm{O(N)}}+\mathcal{O}(1/N^{2})\,,
\label{Fres1b}
\end{align}
where $F_{s}$ and $F_{f}$ are sphere free energies for  conformally coupled free boson and free massless fermion in arbitrary dimensions  \cite{Giombi:2014xxa} 
\begin{align}
F_{s} &=-\frac{1}{\sin(\frac{\pi d}{2})\Gamma(1+d)}\int_0^{1} du\, u \sin \pi u\Gamma (d/2+u )\Gamma (d/2-u) \,,  \label{frees}
\end{align}
\begin{align}
 F_{f} &=-
 \frac{1}{\sin(\frac{\pi d}{2})\Gamma(1+d)}\int_0^{1} du\, \cos\Big(\frac{\pi u}{2}\Big)\Gamma\Big(\frac{1+d+u}{2}\Big)\Gamma\Big(\frac{1+d-u}{2}\Big)\,
 \label{freeff}
\end{align}
and $\delta F_{\Delta}$ is given by \cite{Diaz:2007an, Giombi:2014xxa}
\begin{align}
\delta F_{\Delta} &=\frac{1}{2}\log \det s(x,y)^{-2\Delta} =-
\frac{1}{\sin(\frac{\pi d}{2})\Gamma(1+d)}\int\limits_0^{\Delta-\frac{d}{2}} du\, u\sin\pi u\, \Gamma\Big(\frac{d}{2}+u\Big)\Gamma\Big(\frac{d}{2}-u\Big) \,, \label{Finduce}
\end{align}
where $s(x,y)$ is the chordal distance on $S^{d}$.
Finally  $\gamma_{\phi,1}^{\textrm{O(N)}}$ and $\gamma_{\psi, 1}^{\textrm{GN}}, \gamma_{\psi, 1}^{\textrm{NJL\,}U(1)}, \gamma_{\psi, 1}^{\textrm{NJL\,}SU(2)}$ are $1/N$ corrections to the anomalous dimensions of $\phi$ and $\psi$ fields \cite{Gracey:1990wi,Gracey:1992cp,  Gracey:1993kb}
\begin{align}
\gamma_{\phi, 1}^{\textrm{O(N)}} = \frac{2 \sin (\pi d/2) \Gamma (d-2)}{\pi  \Gamma (d/2-2) \Gamma (d/2+1)}\,, \quad \gamma_{\psi, 1}^{\textrm{GN}} = \frac{(d-2)^2 \Gamma (d-1)}{4 \Gamma \left(2-d/2\right) \Gamma \left(d/2+1\right) \Gamma \left(d/2\right)^2}\,  \label{anomdims}
\end{align}
and $\gamma_{\psi, 1}^{\textrm{NJL\,}U(1)}=\gamma_{\psi, 1}^{\textrm{NJL\,}SU(2)} = 2 \gamma_{\psi, 1}^{\textrm{GN}}$.
In $d=3$ using  (\ref{Fres1}) and (\ref{anomdims}) we find for the $1/N$ term in the sphere free energy\footnote{In $d=3$ we  have $F_{f}=\frac{\log 2}{8}+\frac{3 \zeta (3)}{16\pi ^2}$, $F_{s}=\frac{\log 2}{8}-\frac{3 \zeta (3)}{16\pi ^2}$ and $\delta F_{\Delta =d-1}=-\delta F_{\Delta= d-2}=\frac{\zeta (3)}{8 \pi ^2}$.} 
\begin{align}
F_{1/N}^{O(N)}|_{d=3} =F_{1/N}^{\textrm{GN}}|_{d=3} =F_{1/N}^{\textrm{NJL\,} U(1)}|_{d=3} =-F_{1/N}^{\textrm{NJL\,}SU(2)}|_{d=3} =  \frac{1}{N} \frac{4}{9 \pi ^2}\,.
\label{FGNres1ind=3f}
\end{align}
We notice that the sphere free energies of the Gross-Neveu and $O(N)$ scalar models are related by $\zeta(3) \rightarrow - \zeta(3)$.
One can compare the results (\ref{Fres1}) and (\ref{Fres1b}) in $d=3$ with  the Pade resummed $\eps$-expansion   given in \cite{Fei:2015oha,Fei:2016sgs}. The plots for the GN and NJL $U(1)$ models are depicted in figure \ref{OneNvsEps}.  The plot for the $O(N)$ scalar model is in figure \ref{OneNvsEpsON}. We see that the large $N$ expansion works quite well even for small $N$. Below we obtain the results (\ref{Fres1}) and  (\ref{Fres1b}).
Also we compute different $\eps$-expansions for the fermionic and bosonic theories, which give a cross-check for the large $N$ results. 

The paper is organized as follows: in  section 2 we study the fermionic theories. We start with a supersymmetric example, where it is possible to obtain exact result for the sphere free energy.  Next we consider the Gross-Neveu-Yukawa model at arbitrary even dimension and 
 calculate the sphere free energy for it using the $\eps$-expansion. Finally, we use these $\epsilon$-expansion results to conjecture the $1/N$ correction to the sphere free energy of the Gross-
Neveu model as a function of $d$. Section $3$ is devoted to the bosonic theories.  Namely, we compute the sphere free energy of the six dimensional cubic and  eight dimensional 
 $O(N)$ scalar models using the $\eps$-expansion.  Then we use these $\epsilon$-expansion results to conjecture the $1/N$ correction to the sphere free energy of the $O(N)$ scalar model as a function of $d$. At the end of the paper there are two Appendices, which discuss some attempts towards the analytic  $1/N$ computation.

\newpage
\section{Towards the large $N$ expansion of the sphere free energy in fermionic theory}
\subsection{Supersymmetric example with $O(N)$ symmetry}
An important property of the large $N$ expansion is that all results  as functions of dimension $d$
have a specific structure.  In this section we show this using as an example   a supersymmetric model with $N+1$
chiral superfields and $O(N)$ symmetric superpotential \cite{Nishioka:2013gza, Giombi:2014xxa}
\begin{align}
W = \frac{\lambda}{2} X \sum_{i=1}^{N}Z^{i}Z^{i}\,,
\end{align}
where $X$ and $Z^{i}$ are chiral superfields. 
As was shown in  \cite{Giombi:2014xxa} the exact result for the sphere free energy in arbitrary $d$ is given by
\begin{align}
F = N \mathcal{F}(\Delta_{Z}) + \mathcal{F}(\Delta_{X})\,, \label{SUSYfree}
\end{align}
where $\Delta_{Z}$ and $\Delta_{X}$ are the anomalous dimensions of the chiral superfields $Z$ and  $X$ and 
\begin{align}
\mathcal{F}(\Delta)=2(F_{s}+F_{f}) -\int_{d/2-1}^{\Delta} du \frac{\Gamma(d-1-u)\Gamma(u)\sin(\pi(u-d/2))}{\Gamma(d-1)\sin(\pi d/2)}\,,
\end{align}
where $F_{s}$ and $F_{f}$ are given in (\ref{frees}) and (\ref{freeff}). At the IR fixed point the conformal dimensions are constrained $2\Delta_{Z}+\Delta_{X}=d-1$ and  the value of $\Delta_{Z}$ can be determined by extremizing $F$. Setting $\Delta_{Z}=\frac{d}{2}-1+\gamma_{Z,1}/N+\mathcal{O}(1/N^{2})$ and solving perturbatively in $1/N$ the equation $dF/d\Delta_{Z}=0$ one finds \cite{Giombi:2014xxa}
\begin{align}
\gamma_{Z,1} = -\frac{2 \sin (\frac{\pi  d}{2}) \Gamma (d-2)}{\pi  \Gamma (\frac{d}{2}-1) \Gamma (\frac{d}{2})}\,, \quad \gamma_{Z,2}= 2\gamma_{Z,1}^{2}\Big(\Psi(d)+\frac{1}{d-2}\Big)\,, \label{Zandim}
\end{align}
where  $\Psi(d)=\psi (d-2)-\psi \left(\frac{d}{2}-1\right)+\psi \left(2-\frac{d}{2}\right)-\psi (1)$ and $\psi(x)=\Gamma'(x)/\Gamma(x)$. Now substituting  (\ref{Zandim}) in (\ref{SUSYfree}) and expanding in $1/N$ we find for the $1/N$ and $1/N^{2}$ corrections to the sphere free energy for general $d$
\begin{align}
F_{1/N}=\frac{1}{N}\frac{\gamma_{Z,1}}{d-2}\,, \quad F_{1/N^{2}}=\frac{1}{N^{2}}\gamma_{Z,1}^{2}\left(\frac{2 \Psi (d)}{d-2}+\frac{4}{3 (d-2)^2}\right)\,. \label{SUSYres2}
\end{align}
We notice that the $1/N$ terms in (\ref{Fres1}) and (\ref{Fres1b})  have the same structure, as in (\ref{SUSYres2}).

\subsection{Gross-Neveu-Yukawa model in general even dimension}
\label{secGNY}

\begin{figure}[h!]
                \centering
                \includegraphics[width=8cm]{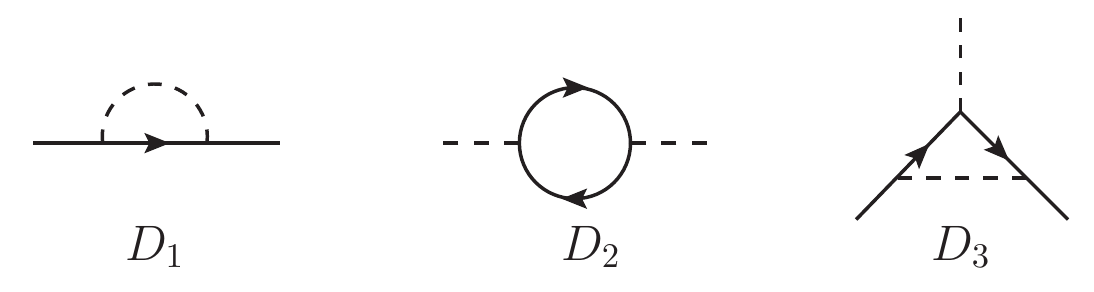}
                 \caption{Feynman diagrams contributing to the anomalous dimension of $\psi$ and $\sigma$ fields and  the beta-function $\beta_{g_{1}}$ at the leading order in $\eps$. The dashed lines represent $\sigma$ field propagators $\langle \sigma \sigma \rangle = 1/(p^{2})^{k-1}$ and the solid lines are for $\psi$ field propagators $\langle \psi \bar{\psi}\rangle = i \slashed{p}/p^{2}$.}
\label{GNevenDdiags}
\end{figure}

In this section we consider the Gross-Neveu-Yukawa model in general even dimension $d=2k$, where $k=2,3,4,\dots$. The action for this model in flat $d=2k-\eps$ dimensional space reads
\begin{align}
S_{\textrm{GNY}}= \int d^{d}x  \Big(\bar{\psi}_{i}\slashed{\partial}\psi^{i}+  (\partial^{k-1}\sigma)^{2} +\mu^{\eps/2} g_{1}\sigma \bar{\psi}_{i}\psi^{i}+\dots\Big)\,, \label{genGNact}
\end{align}
where $\psi_{i}$, $i=1,\dots, N_{f}$ are Dirac fermions,     $\sigma$ is a higher-derivative scalar field and $\mu$ is an auxiliary scale. Dots in (\ref{genGNact}) denote interacting terms made from $\sigma$ field (and its derivatives), and also all counter-terms. For $k=2$ the GNY model was proposed in \cite{Hasenfratz:1991it,ZinnJustin:1991yn}  as a UV completion of the GN model. The sphere free energy for the $k=2$ GNY model was computed recently in \cite{Fei:2016sgs}.     For general $k$ the bare dimensions of $\psi$ and $\sigma$ fields are $[\psi]= [\mu]^{\frac{d-1}{2}}$ and $[\sigma]= [\mu]^{\frac{d+2-2k}{2}}$\,. At the leading order in $\eps$  only $g_{1}$-vertex contributes to the anomalous dimension of $\psi$ and $\sigma$ fields and to the beta-function $\beta_{g_{1}}$. To compute them, we need to consider three simple one-loop diagrams, depicted in  figure \ref{GNevenDdiags}.

\noindent Working in the MS scheme, it is not difficult to compute the counter-terms $\delta_{\psi}$, $\delta_{\sigma}$ and $\delta_{g_{1}}$, defined as  $\psi = Z_{\psi}^{-1/2}\psi_{0}$, $\sigma = Z_{\sigma}^{-1/2}\sigma_{0}$, where $Z_{\psi} = 1+\delta_{\psi}$ and $Z_{\sigma}=1+\delta_{\sigma}$ and also $g_{1,0} =Z_{g} Z_{\psi}^{-1}Z_{\sigma}^{-1/2} g$, where $Z_{g}= 1+\delta_{g_{1}}/g_{1}$. The result is 
 \begin{align}
 \delta_{\psi}=-\frac{ 2^{1-2 k} (k-1) }{\pi ^{k}  \Gamma (k+1)} \frac{g_{1}^{2}}{\epsilon}\,, \quad \delta_{\sigma}=N\frac{ (-1)^{k+1} 2^{3-4 k}  }{ \pi ^{k-\frac{1}{2}} \Gamma \left(k-\frac{1}{2}\right)} \frac{g_{1}^{2}}{\eps}\,, \quad \delta_{g_{1}}=\frac{ 2^{1-2 k} }{\pi ^{k}  \Gamma (k)} \frac{g_{1}^{3}}{\eps}\,,
 \end{align}
 where $N=N_{f} \Tr \bf{1}$.   Therefore for the anomalous dimensions and beta-function we find 
  \begin{align}
 &\gamma_{\psi}=\frac{ k-1}{(4\pi)^{k}  \Gamma (k+1)} g_{1}^{2}+\dots\,, \quad  \gamma_{\sigma}=-\frac{ N(-1)^{k+1}   }{ (16\pi )^{k-\frac{1}{2}} \Gamma \left(k-\frac{1}{2}\right)} g_{1}^{2} +\dots\,, \notag\\
&\beta_{g_{1}}=-\frac{\eps}{2}g_{1}+  \left(\frac{2 \sqrt{\pi } (-1)^k N}{\Gamma \left(k-\frac{1}{2}\right)}+\frac{4^k (2 k-1)}{\Gamma (k+1)}\right)\frac{g_{1}^{3}}{ 2^{4 k-1}\pi ^{k}} +\dots\,, \label{gbetares}
 \end{align}
 where the full dimensions of $\psi$ and $\sigma$ fields are $\Delta_{\psi} = \frac{d-1}{2}+\gamma_{\psi}$ and $\Delta_{\sigma}=\frac{d+2-2k}{2}+\gamma_{\sigma}$\,. Thus from (\ref{gbetares}) the  fixed point at the leading order in $\eps$ is 
\begin{align}
g^{2}_{1,*} = 4^{2k-1}\pi^{k}   \bigg(\frac{2 \sqrt{\pi } (-1)^k N}{\Gamma \left(k-\frac{1}{2}\right)}+\frac{4^k (2 k-1)}{\Gamma (k+1)}\bigg)^{-1}\eps+\mathcal{O}(\eps^{2})\,. \label{gGNevencrit}
\end{align}
And we find for the anomalous dimensions at the critical point  \footnote{These results were obtained  in collaboration with Lin Fei. } 
\begin{align}
&\gamma_{\psi}=\frac{(k-1)^2 k   \Gamma (2 k-2)}{N(-1)^k  \Gamma (k+1)^2+\Gamma (2 k+1)}\epsilon+\mathcal{O}(\eps^{2})\,,\notag\\
&\gamma_{\sigma}=\frac{N(-1)^k \sqrt{\pi }  \Gamma (k+1)}{2 \big(N (-1)^k\sqrt{\pi }  \Gamma (k+1)+4^k \Gamma (k+\frac{1}{2})\big)}\epsilon+\mathcal{O}(\eps^{2})\,. \label{gammaeps}
 \end{align}
One can compare these results to the anomalous dimensions of $\psi$ and $\sigma$ fields computed in the  Gross-Neveu model in the large $N$ approximation.  We have $\Delta_{\psi} = \frac{d-1}{2}+\gamma_{\psi,1}/N+\gamma_{\psi,2}/N^{2}+\mathcal{O}(1/N^{3})$ and $\Delta_{\sigma}= 1+\gamma_{\sigma,1}/N+\mathcal{O}(1/N^{2})$, where the functions $\gamma_{\psi,1}, \gamma_{\psi,2}, \gamma_{\psi,3}$ and $\gamma_{\sigma,1}$, $\gamma_{\sigma,2}$ are given in \cite{Gracey:1990wi,Gracey:1992cp,  Gracey:1993kb, Gracey:1993kc, Vasiliev:1992wr}.
Expanding these functions in $\eps$  at  $d=2k-\eps$ one finds perfect agreement with    (\ref{gammaeps}).
 
 Next we  compute the critical free energy on the sphere for the model (\ref{genGNact}) at  $d=2k-\eps$ for an arbitrary positive integer $k$.  At the leading order in $\eps$ only the $g_{1}$-vertex contributes and 
 we have
  \begin{align}
F_{\textrm{GNY}}=NF_{f}+F_{\sigma}-\frac{1}{2!}g_{1,*}^{2} \int d^{d}xd^{d}y\sqrt{g_{x}}\sqrt{g_{y}}\langle\sigma \bar{\psi}_{i}\psi^{i}(x) \sigma \bar{\psi}_{j}\psi^{j}(y)\rangle+\mathcal{O}(\eps^{3})\,, \label{dFGN1}
 \end{align}
 where $F_{f}$ is the sphere free energy for the free massless Dirac fermion (\ref{freeff})  and $F_{\sigma}$ is the sphere free energy for 
 the free higher-derivative scalar $\sigma$, $F_{\sigma}$ is given by (\ref{Finduce}) with $\Delta=\frac{d}{2}+k-1$.
The propagators of $\psi$ and $\sigma$ fields in  flat coordinate space are
   \begin{align}
\langle \psi_{i}(x)\bar{\psi}^{j}(0)\rangle=-\delta_{i}^{j}\frac{\Gamma \left(\frac{d}{2}\right)}{2 \pi ^{d/2}} \frac{\gamma_{\mu}x_{\mu}}{(x^{2})^{\frac{d}{2}}}\,, \quad \langle \sigma(x) \sigma(0) \rangle =\frac{\ \Gamma \left(\frac{d}{2}-k+1\right)}{4^{k-1}\pi^{\frac{d}{2}} \Gamma (k-1)} \frac{1}{(x^{2})^{\frac{d}{2}+1-k}}\,.
 \end{align}
 Because the propagators in the flat space are related to the sphere ones by the Weyl transformation,
 we can compute the correlation function in (\ref{dFGN1}) working in the flat space and then project the result on the sphere by making replacement $(x-y)^{2}\to s(x,y)^{2}$, where $s(x,y)$ is the chordal distance on the sphere. 
We find 
   \begin{align}
F_{\textrm{GNY}}=NF_{f}+F_{\sigma}-\frac{1}{2!}g_{1,*}^{2}N \Big(\frac{\Gamma \left(\frac{d}{2}\right)}{2 \pi ^{d/2}}\Big)^{2}\frac{\ \Gamma \left(\frac{d}{2}-k+1\right)}{4^{k-1}\pi^{\frac{d}{2}} \Gamma (k-1)}I_{2}\big(\frac{3d}{2}-k\big) +\mathcal{O}(\eps^{3})\,, \label{dFGN2}
 \end{align}
 where the integral $I_{2}(\Delta)$ is defined as \cite{Drummond:1977dn, Cardy:1988cwa, Klebanov:2011gs}
\begin{equation}
I_{2}(\Delta)=\int \frac{d^{d}xd^{d}y\sqrt{g_{x}}\sqrt{g_{y}}}{s(x,y)^{2\Delta}}= (2 R)^{2 (d-\Delta )}\frac{2^{1-d} \pi ^{d+\frac{1}{2}} \Gamma \left(\frac{d}{2}-\Delta \right)}{\Gamma \left(\frac{d+1}{2}\right) \Gamma (d-\Delta )}\, 
\label{I2}
\end{equation}
and $R$ is the sphere radius.  Now substituting  (\ref{gGNevencrit})  in (\ref{dFGN2}) and expanding everything in $\eps$ at $d=2k-\eps$ we find in terms of 
generalized free energy $\tilde{F}= -\sin(\pi d/2)F$, which is a smooth function of dimension \cite{Giombi:2014xxa}
   \begin{align}
\delta \tilde{F}_{\textrm{GNY}}=  -\frac{N \pi ^{3/2} k   \Gamma (k)^2 \Gamma (2 k-1)}{8 \Gamma (k-1) \Gamma (2 k+1) \left(N\sqrt{\pi } (-1)^k k \Gamma (k)+4^k \Gamma (k+\frac{1}{2})\right)} \epsilon ^2 +\mathcal{O}(\eps^{3})\,, \label{dFGN3}
 \end{align}
 where we denoted $\delta \tilde{F}_{\textrm{GNY}}=\tilde{F}_{\textrm{GNY}}-N\tilde{F}_{f}-\tilde{F}_{\sigma}$.
Expanding (\ref{dFGN3}) in the large $N$ limit we get 
\begin{align}
\delta \tilde{F}_{\textrm{GNY}}= \bigg(-\frac{\pi  (-1)^k (k-1)}{16 k (2 k-1)}+\frac{1}{N}\frac{\sqrt{\pi } 2^{2 k-5} \Gamma \left(k-\frac{1}{2}\right)}{k^2 \Gamma (k-1)}+\mathcal{O}(1/N^{2})\bigg)\epsilon ^2+\mathcal{O}(\eps^{3})\,. \label{dFGNres}
\end{align}
One can check that the $\eps^2/N$ term obtained by expansion of  (\ref{Fres1}) in small $\eps$ near $d=2k-\eps$ coincides with the same term in  (\ref{dFGNres})

\subsection{The $1/N$ correction to the GN model}
\label{GNlargNfree}
Based on analytical computations described in Appendix A and analogy with the supersymmetric result (\ref{SUSYres2}) we may guess the expression for the $1/N$ term in the GN sphere free energy
\begin{align}
\tilde{F}^{\textrm{GN}}_{1/N}= -\frac{\sin(\pi d/2)}{N d}\gamma_{\psi, 1}^{\textrm{GN}}\,,
\label{FGNguess}
\end{align}
where  $\gamma_{\psi, 1}^{\textrm{GN}}
$ is given in (\ref{anomdims}).
One can verify  that (\ref{FGNguess}) gives the correct $\eps$ expansion near various integer dimensions. 
Let us expand it
 in small $\eps$ near $d=2$ and $d=4$. We find
\begin{align}
 \tilde{F}^{\textrm{GN}}_{1/N}|_{d=2+\eps}=\frac{1}{N}\Big(\frac{\pi  \epsilon ^3}{16}-\frac{\pi  \epsilon ^4}{16}+\mathcal{O}(\eps^5) \Big)\,, \quad 
 \tilde{F}^{\textrm{GN}}_{1/N}|_{d=4-\eps}=\frac{1}{N}\Big(\frac{\pi  \epsilon ^2}{16}-\frac{\pi  \epsilon ^3}{32}+\mathcal{O}(\eps^4) \Big)\,.
\end{align}
We have checked that these expansions match with the $\eps$-expansions in  $d=2+\eps$ and $d=4-\eps$ for the sphere free energy of the Gross-Neveu and Gross-Neveu-Yukawa models 
obtained in \cite{Fei:2016sgs}. Moreover, as we have already explained, the $1/N$ term in (\ref{FGNguess}) matches with the result (\ref{dFGNres}) for expansion at $d=2k-\eps$ for  all $k=2,3,4,\dots$. 

\noindent The $\eps$-expansion for the NJL $U(1)$  model  near $d=2$ and $4$  coincides with the result in \cite{Fei:2016sgs}.



\section{Towards the large $N$ expansion of sphere free energy in bosonic theory}
\subsection{Free energy on $S^{6-\eps}$ for cubic $O(N)$ scalar theory}
\label{cubicONsec}
In this section we consider the cubic $O(N)$ model on a sphere in $d=6-\eps$ dimension. This model in flat space was  recently proposed  
in \cite{Fei:2014yja}  and further studied in \cite{Fei:2014xta, Fei:2015kta, Gracey:2015tta} (non-perturbative studies of this fixed point have been carried out using functional renormalization group \cite{Mati:2014xma, Mati:2016wjn, Eichhorn:2016hdi, Kamikado:2016dvw }
and conformal bootstrap \cite{Chester:2014gqa,Li:2016wdp }). The leading term in $\eps$-expansion for the sphere free energy was already computed in \cite{Giombi:2014xxa}.  Here we calculate the next to leading term.
In order to renormalize the theory on a curved manifold, one should add to the action all the relevant curvature couplings that are marginal 
in $d=6-\eps$ \cite{Brown:1980qq,Hathrell:1981zb}
\begin{align}
S_{\textrm{cubic}\, O(N)}=& \int d^{d}x \sqrt{g} \Big(\frac{1}{2}(\partial_{\mu}\phi^{i})^{2}+\frac{1}{2}(\partial_{\mu}\sigma)^{2}+\frac{d-2}{8(d-1)} \mathcal{R}(\sigma^{2}+\phi^{i}\phi^{i})+\frac{1}{2}g_{1,0}\sigma \phi^{i}\phi^{i}+\frac{1}{6}g_{2,0}\sigma^{3}\Big)\notag\\
&+\frac{\eta_0}{2}{\cal R}(\sigma^2+\phi^{i}\phi^{i})+ b_{0}\int d^{d}x \sqrt{g} \mathcal{R}^{3}\,,
\end{align}
where $\sigma$ and $\phi^{i}$, $i=1,\dots, N$ are  bare scalar fields, $g_{1,0}$ and $g_{2,0}$ are bare  couplings and ${\cal R}$ is the scalar curvature. 
The parameters $\eta_0, b_0$ are bare curvature couplings whose renormalization is fixed order by order in perturbation 
theory. Below we will only need the renormalization of 
the coupling $b_0$ (the renormalization of conformal coupling $\eta_{0}$ is expected to play a role at higher orders 
\cite{Brown:1980qq,Hathrell:1981zb,Jack:1990eb}).  We will find the beta function for $b_{0}$ coupling by demanding that 
the final result for the free energy is free of divergences.

\begin{figure}[h!]
                \centering
                \includegraphics[width=14cm]{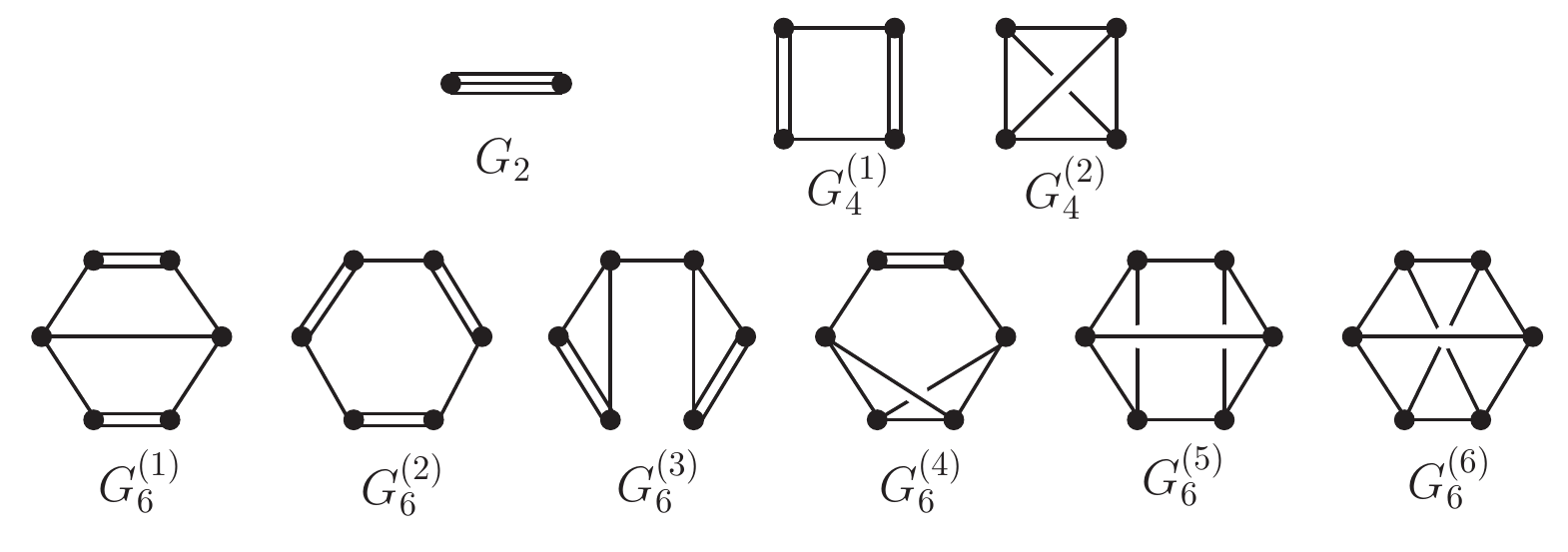}
                 \caption{Diagrams contributing to the sphere free energy in the cubic $O(N)$ theory up to the sixth order. Each line represents the propagator $\langle \phi(x)\phi(y)\rangle = C_{\phi}/s(x,y)^{d-2}$.}
\label{diagsPhi3}
\end{figure}

It is convenient to combine the cubic interaction terms in the  form 
\begin{align}
\mathcal{L}_{\textrm{int}} = \frac{1}{3!}g_{abc}\varphi^{a}\varphi^{b}\varphi^{c}\,,
\end{align}
with symmetric tensor $g_{abc}$, where $a,b,c=1,\dots ,N+1$,
and we identify $\varphi^{0}=\sigma$ and $\varphi^{i}=\phi^{i}$ and  $g_{000}=g_{2,0}$,  $g_{0ii}=g_{i0i}=g_{ii0}=g_{1,0}$ and all other components of the tensor $g_{abc}$ are zero. 
Next we define integrated connected correlation functions on a sphere
\begin{align}
G_{n}= \int \prod_{i=1}^{n} dx_{i}\sqrt{g_{x_{i}}}\langle \varphi^{3}(x_{1})\dots \varphi^{3}(x_{n})\rangle_{0}^{\textrm{conn}}\,,
\end{align}
where $\varphi^{3}\equiv g_{abc}\varphi^{a}\varphi^{b}\varphi^{c}$ and the two-point function on the sphere is 
 \begin{align}
\langle \varphi^{a}(x)\varphi^{b}(y)\rangle = \delta^{ab}\frac{\Gamma(\frac{d}{2}-1)}{4\pi^{d/2}} \frac{1}{s(x,y)^{d-2}}= \delta^{ab} \frac{C_{\phi}}{s(x,y)^{d-2}}\,. \label{phiprop}
\end{align}
Evidently $G_{n}=0$ if $n$ is odd. Therefore for the free energy on the sphere up to the sixth order in couplings  we find
\begin{align}
F_{\textrm{cubic}\, O(N)}= (N+1)F_{s}-\frac{1}{2!}\frac{G_{2}}{(3!)^{2}}-\frac{1}{4!}\frac{G_{4}}{(3!)^{4}}
 -\frac{1}{6!}\frac{G_{6}}{(3!)^{6}}+ b_{0}\int d^{d}x \sqrt{g} \mathcal{R}^{3}\,, \label{Fcub1}
\end{align}
where 
\begin{align}
&G_{2} =3!t_{2} C_{\phi}^{3}I_{2}\big(\frac{3}{2}(d-2)\big)\,, \notag\\
&G_{4}=3(3!)^{3}(3t_{41}G_{4}^{(1)}+2t_{42}G_{4}^{(2)})\,, \notag\\
&G_{6}=15(3!)^{5}(18 t_{61} G_{6}^{(1)}+6 t_{62}G_{6}^{(2)}+9 t_{63}G_{6}^{(3)}+36 t_{64}G_{6}^{(4)}+24t_{65} G_{6}^{(5)}+4t_{66}G_{6}^{(6)})\,, \label{corYL}
\end{align}
where   the integral $I_{2}(\Delta)$ is defined in (\ref{I2}) and  $t$-coefficients are computed by contracting the  tensors  $g_{abc}$ for each diagram depicted in figure  \ref{diagsPhi3}\,\footnote{These results were obtained in collaboration with Lin Fei. } 
\begin{align}
t_{2}&=3 N g_{1}^{2}+g_{2}^2\,,\quad  t_{41}=(N+4)Ng_{1}^{4} +2 N g_{1}^{2} g_{2}^{2}+g_{2}^{4}\,, \quad t_{42}=3N g_{1}^{4} +4 N g_{1}^{3} g_{2} +g_{2}^{4}\,, \notag\\
t_{61}&=4N (N+1) g_{1}^{6} +N (N+4)g_{1}^{4} g_{2}^{2} +2N g_{1}^2 g_{2}^{4} +g_{2}^6\,,\quad t_{62}=8N g_{1}^6 +(Ng_{1}^2 +g_{2}^2)^3\,, \notag \\
t_{63}&=N^2 g_{1}^4 (2 g_{1}+g_{2})^2+4 N g_{1}^3 g_{2}^3 +g_{2}^6\,, \notag\\
t_{64}&=N (N+4)g_{1}^6 +2  N(N+2)g_{1}^5 g_{2}+Ng_{1}^4 g_{2}^2 +2N g_{1}^3 g_{2}^3 +N g_{1}^2 g_{2}^4 +g_{2}^6\,, \notag\\
t_{65}&=N(N+3)g_{1}^6 +6 N g_{1}^5 g_{2} +3 N g_{1}^4 g_{2}^2 +2 Ng_{1}^3 g_{2}^3 +g_{2}^6\,, \quad t_{66}=6N g_{1}^6 +9N g_{1}^4 g_{2}^2 +g_{2}^6\,.
\end{align}
\noindent Here for brevity we suppressed  the index $_{0}$ indicating that these couplings are bare.

\noindent To compute higher-loop integrals  we use the Mellin-Barnes approach, which is described in \cite{Fei:2015oha, Giombi:2015haa}. We find for the four-point functions
\begin{align}
G_{4}^{(k)}= C_{\phi}^{6} (2 R)^{2 (6-d)}\frac{2^{1-d} \pi ^{\frac{d+1}{2}}}{\Gamma \left(\frac{d+1}{2}\right)} e^{\frac{3}{2}(d-6)\gamma}\pi^{\frac{3d}{2}}
\begin{cases}
\frac{1}{18 \epsilon }+\frac{43}{162}+\big(\frac{2857}{3888}+\frac{\pi ^2}{288}\big) \epsilon +\mathcal{O}(\epsilon^{2})\,, \\
 -\frac{1}{3 \epsilon }-\frac{3}{2}-\big(\frac{191}{48}+\frac{\pi ^2}{48}\big) \epsilon+\mathcal{O}(\epsilon^{2})\,,
\end{cases} \label{G4res}
\end{align}
where $R$ is the sphere  radius and also for the six-point functions we obtain
\begin{align}
G_{6}^{(k)}=C_{\phi}^{9} (2 R)^{3 (6-d)}\frac{2^{1-d} \pi ^{\frac{d+1}{2}}}{\Gamma \left(\frac{d+1}{2}\right)} e^{\frac{5}{2}(d-6)\gamma}\pi^{\frac{5d}{2}}
\begin{cases}
-\frac{1}{54 \epsilon ^2}-\frac{389}{2592 \epsilon }-\big(\frac{87}{128}+\frac{\pi ^2}{864}\big)+\mathcal{O}(\epsilon)\,,\\
-\frac{1}{36 \epsilon ^2}-\frac{139}{648 \epsilon }-\big(\frac{3547}{3888}+\frac{\pi ^2}{576}\big)+\mathcal{O}(\epsilon)\,,\\
\mathcal{O}(\epsilon)\,,\\
\frac{1}{9 \epsilon ^2}+\frac{733}{864 \epsilon }+\big(\frac{4675}{1296}+\frac{\pi ^2}{144}\big)+\mathcal{O}(\epsilon)\,, \\
-\frac{1}{3 \epsilon ^2}-\frac{703}{288 \epsilon }-\big(\frac{34649}{3456}+\frac{\pi ^2}{48}\big)+\mathcal{O}(\epsilon)\,,\\
-\frac{5}{24 \epsilon }+\big(\frac{\zeta (3)}{8}-\frac{475}{288}\big)+\mathcal{O}(\epsilon)\,.
\end{cases} \label{G6res}
\end{align}
The beta-functions for the cubic $O(N)$ model were computed in \cite{Fei:2014yja, Fei:2014xta, Gracey:2015tta}. For our calculation we need to extract the relation between the bare couplings and the renormalized ones 
\begin{align}
&g_{0,1}=\mu^{\epsilon/2}\bigg(g_{1}+\frac{(N-8)g_{1}^{3}-12g_{1}^{2}g_{2}+g_{1}g_{2}^{2}}{12(4\pi)^{3}} \frac{1}{\epsilon}+\dots\bigg)\,, \notag\\
&g_{0,2}=\mu^{\epsilon/2}\bigg(g_{2}- \frac{3g_{2}^{3}+4N g_{1}^{3}-N g_{1}^{2}g_{2}}{4(4\pi)^{3}} \frac{1}{\epsilon}+\dots\bigg)\,, 
\label{cubicgbgr}
\end{align}
where $\mu$ is the auxiliary scale and we have not  explicitly written higher order terms.  Combining all the results (\ref{corYL}), (\ref{G4res}), (\ref{G6res}) in (\ref{Fcub1}) and using (\ref{cubicgbgr}), we find that in order to cancel all poles in $\eps$ we should renormalize $b_{0}$ coupling as
\begin{align}
&b_{0}=\mu ^{-\epsilon } \left(b +\frac{N+1}{756\cdot 450 (4 \pi )^{3}}\frac{1}{\eps} +\frac{b_{61}(g_{1},g_{2})}{\epsilon }+\dots\right)\,, \label{b0bb}
\end{align}
where 
\begin{align}
b_{61}(g_{1},g_{2})=&\frac{1}{2^{12} 3^8 5^3 (4 \pi )^{12}} \Big(N \big(2  (43 N+268)g_{1}^6-12 (11 N-32)g_{1}^5 g_{2} \notag\\
&~~~+(11N+950)g_{1}^4 g_{2}^2 +84 g_{1}^3 g_{2}^3-44 g_{1}^2 g_{2}^4\big)+125 g_{2}^6\Big)\,
\end{align}
and the first coupling independent term in (\ref{b0bb}) is due to the trace anomaly of the free  boson field.
Thus the beta-function for the $b$ coupling reads 
\begin{align}
\beta_{b} = \epsilon b + \frac{N+1}{756\cdot 450 (4 \pi )^{3}} +4 b_{61}(g_{1},g_{2})\label{Cubbetab}\,.
\end{align}
As explained in \cite{Fei:2015oha}, in order to calculate the free energy at the critical point we should  tune all couplings, including $b$, to their 
fixed point values. Using (\ref{Cubbetab}) and the results from \cite{Fei:2014xta} for the large $N$ critical point of the $O(N)$ cubic model, we find
\begin{align}
&g^{*}_{1} = \sqrt{\frac{6\eps(4\pi)^3}{N}}\left(1 + \frac{22}{N}+\frac{726}{N^2}-\left( \frac{155}{6N} + \frac{1705}{N^2} \right)\eps +  \frac{1777}{144N}\eps^2+\dots\right)\,, \notag\\
& g_{2}^{*} =6 \sqrt{\frac{6\eps(4\pi)^3}{N}}\left( 1 + \frac{162}{N}+\frac{68766}{N^2}- \left( \frac{215}{2N}+\frac{86335}{N^2}\right)\eps + \frac{2781}{48N}\eps^2+\dots\right)\,, \notag\\
&b_{*}= -\frac{N+1}{756\cdot 450 (4 \pi )^{3}}\frac{1}{\eps} - \frac{4b_{61}(g^{*}_{1},g^{*}_{2})}{\eps}+\dots\,, \label{critvalcub}
\end{align}
where the dots stand for higher order terms in $\eps$ and $1/N$.
The contribution to the sphere free energy from the curvature coupling reads\footnote{Note that we computed the six-point functions in (\ref{G6res}) essentially only to find the renormalization of $b$. If one finds the next order  term in the renormalization of $b$, then,  using (\ref{G4res}) and (\ref{G6res}), it is possible to get the next order term for the sphere free energy. } 
\begin{align}
\delta F_{b} = b_{*}\int d^{d}x \sqrt{g} \mathcal{R}^{3} = 450(4\pi)^{3} b_{*} +\mathcal{O}(\eps^{3})\,,
\end{align}
where we used that  $\textrm{vol}(S^{d})=\int d^{d}x \sqrt{g}= 2\pi^{(d+1)/2}R^{d}/\Gamma(\frac{d+1}{2})$ and the scalar curvature on the sphere is $\mathcal{R}= d(d-1)/R^{2}$.  Substituting the values (\ref{critvalcub}) in  (\ref{Fcub1})
we get in terms of $\tilde{F}=-\sin(\pi d/2) F$
\begin{align}
&\tilde{F}_{\textrm{cubic}\, O(N)}=(N+1)\tilde{F}_{s}-\left(\frac{\pi  \epsilon ^2}{960}+\frac{19 \pi  \epsilon ^3}{43200}\right)-\left(\frac{7 \pi  \epsilon ^2}{120}-\frac{17 \pi  \epsilon ^3}{400}\right)\frac{1}{N}
\notag\\
&~~~~~~-\left(\frac{91 \pi  \epsilon ^2}{15}-\frac{56843 \pi  \epsilon ^3}{5400}\right)\frac{1}{N^2}+\mathcal{O}\Big(\eps^{4},\frac{1}{N^{3}}\Big)\,. \label{FcubONres}
\end{align}
Expanding the large $N$ result for the $O(N)$ scalar model (\ref{Fres1}) in small $\eps$ at $d=6-\eps$ we find perfect agreement with the $1/N$ term in (\ref{FcubONres}).

\subsection{Free energy on $S^{8-\eps}$ for $O(N)$ scalar theory}

In this section we compute the leading  in $\eps$ term for the sphere free energy of  the $O(N)$ scalar model at $d=8-\eps$ dimensions.
At  leading order, one need  not to take into account contributions to the free energy coming from the curvature
couplings.  The curvature couplings contribute only at the next to leading order in $\eps$, as we saw in the previous section.

The action for the $O(N)$ model in $d=8-\eps$ flat space reads \cite{Gracey:2015xmw}
\begin{align}
S_{8d\; O(N)} = \int d^{d}x \Big(\frac{1}{2}(\partial_{\mu}\phi^{i})^{2}+\frac{1}{2}(\Box \sigma)^{2}+\frac{1}{2}\mu^{\frac{\eps}{2}}g_{1}\sigma \phi^{i}\phi^{i}+\frac{1}{6}\mu^{\frac{\eps}{2}}g_{2}\sigma^{2}\Box \sigma +\frac{1}{4!}\mu^{\eps}g_{3}^{2}\sigma^{4}+\dots\Big)\,, \label{d8ONact}
\end{align}
where $\sigma$ and $\phi^{i}$, $i=1,\dots,N$ are  scalar fields, $g_{1}, g_{2}$ and $g_{3}$ are renormalized  couplings and $\mu$ is an auxiliary scale. Dots in (\ref{d8ONact}) stand for the counter-terms. The beta-functions and anomalous dimensions for this model  were computed up to three loops in \cite{Gracey:2015xmw}. 

At leading order in $\eps$ only terms with $g_{1}$ and $g_{2}$ contribute to the sphere free energy. 
The contribution, which involves  $g_{2}$ coupling, requires a careful treatment.  Let us introduce two operators
\begin{align}
O_{1}(x) = \sigma^{2}\Box \sigma, \quad O_{2}(x)= \sigma (\partial \sigma)^{2}\,.
\end{align}
First, an operator $\mathcal{D}(x)= O_{1}(x)+2O_{2}(x)$, being a total derivative, is a descendant operator. Then we can find the primary operator  $\mathcal{P}(x)$ as a linear combination of the operators $O_{1}$ and $O_{2}$, which is orthogonal to the descendant operator $\langle \mathcal{P}(x)\mathcal{D}(y)\rangle =0$. One finds that the primary operator is given by the formula
\begin{align}
\mathcal{P}(x)=\frac{1}{2}(d-4)O_{1}(x)+O_{2}(x)\,.
\end{align}
Therefore for the operator $O_{1}(x)$, which appears in the Lagrangian (\ref{d8ONact}) we get 
\begin{align}
O_{1}(x) = \frac{1}{d-5}(2\mathcal{P}(x)-\mathcal{D}(x))\,.
\end{align}
Then at the leading order for  the sphere free energy one finds 
\begin{align}
 F_{8d\; O(N) }=&NF_{s}+F_{\sigma}  - \frac{1}{2!}\frac{1}{2^2}g_{1}^{2} \int d^{d}xd^{d}y\sqrt{g_{x}}\sqrt{g_{y}} \langle \sigma \phi^{i}\phi^{i}(x) \sigma \phi^{j}\phi^{j}(y)\rangle \notag\\
&-\frac{1}{2!}\frac{1}{6^{2}}g_{2}^{2} \Big(\frac{2}{d-5}\Big)^{2} \int d^{d}xd^{d}y\sqrt{g_{x}}\sqrt{g_{y}} \langle \mathcal{P}(x) \mathcal{P}(y)\rangle+\dots\,, \label{d8ONfreeEn}
\end{align}
where $F_{s}$ is the sphere free energy for the free conformally coupled scalar  (\ref{frees})  and $F_{\sigma}$ is the sphere free energy for 
 the free higher-derivative scalar $\sigma$,  $F_{\sigma}$ is given by (\ref{Finduce}) with $\Delta=\frac{d}{2}+2$.
The propagator for the $\phi$ field is given in (\ref{phiprop})  and   the  $\sigma$ field propagator reads
\begin{align}
 \langle \sigma(x) \sigma(y)\rangle= \frac{\Gamma \left(\frac{d}{2}-2\right)}{16 \pi ^{d/2}}\frac{1}{s(x,y)^{d-4}}=  \frac{C_{\sigma}}{s(x,y)^{d-4}} \,. \label{sigmaprop}
\end{align}
Using (\ref{sigmaprop}) we obtain for the two-point function of the primary operators
\begin{align}
 \langle \mathcal{P}(x) \mathcal{P}(y)\rangle=  -2d(d-5)(d-4)^{2}\frac{C_{\sigma}^{3}}{s(x,y)^{3d-8}}\,.
\end{align}
Finally computing the correlation functions in  (\ref{d8ONfreeEn})  and tuning the couplings to the critical point we find
for the shift of the sphere free energy $\delta \tilde{F} = -\sin(\pi d/2)(F_{8d \;O(N) }-NF_{s}-F_{\sigma})$ 
\begin{align}
\delta \tilde{F}_{8d\;O(N)}=&  \sin(\pi d/2)\Big(\frac{1}{4}N g_{1,*}^{2}C_{\phi}^{2}C_{\sigma}-\frac{d(d-4)^2  }{9 (d-5)}g_{2,*}^{2}C_{\sigma}^{3}\Big)I_{2}\Big(\frac{3d}{2}-4\Big)+\mathcal{O}(\eps^{3})\,, \label{dF81}
\end{align}
where   the integral $I_{2}(\Delta)$ is defined in (\ref{I2}) and the large $N$ critical couplings were computed in \cite{Gracey:2015xmw}
and read
\begin{align}
&g^{*}_{1}= i\sqrt{\frac{60(4\pi)^{4}\eps}{N}}\left(1+(-110+126 \epsilon )\frac{1}{N}+(18150-\frac{48475}{2} \epsilon)\frac{1}{N^{2}} +\mathcal{O}(1/N^{3})\right)\,,\\
&g^{*}_{2} =  -i\sqrt{\frac{60(4\pi)^{4}\eps}{N}}\left( 15 - (14250 - 10605\eps) \frac{1}{N} + (36182250 - \frac{89882625}{2}\eps)\frac{1}{N^{2}}+\mathcal{O}(1/N^{3})\right)\,. \notag
\end{align}
Expanding the expression (\ref{dF81}) in small $\eps$ and $1/N$ we get
\begin{align}
\delta \tilde{F}_{8d\;O(N)}=\Big(\frac{\pi }{1344}-\frac{1}{N}\frac{215 \pi }{1008 }+\frac{1}{N^{2}}\frac{32825 \pi }{252}+\mathcal{O}(1/N^{3})\Big) \epsilon ^2+\mathcal{O}(\eps^{3})\,.
 \label{dF82res}
\end{align}
Expanding the large $N$ result for the $O(N)$ scalar model (\ref{Fres1}) in small $\eps$ at $d=8-\eps$ we find perfect agreement with the $1/N$ term in (\ref{dF82res}).

\subsection{The $1/N$ correction to the $O(N)$ scalar model }
Based on analytical computations described in Appendix B and analogy with the supersymmetric result (\ref{SUSYres2}) we may guess the expression for the $1/N$ term in the $O(N)$ sphere free energy 
\begin{align}
\tilde{F}^{O(N)}_{1/N}= -\frac{1}{N }\sin\big(\frac{\pi d}{2}\big)\Big(\frac{1}{d}+\frac{1}{3 (d-2)}-\frac{2}{3 (d-1)}\Big)\gamma_{\phi,1}^{\textrm{O(N)}} \label{ONguess}\,,
\end{align}
where $\gamma_{\phi,1}^{\textrm{O(N)}}$ is given in (\ref{anomdims}). 
One can verify that (\ref{ONguess}) gives the correct $\eps$-expansion near various integer dimensions.
Let us expand  it in small $\eps$ near $d=4$, $d=6$ and $d=8$ dimensions. We find
\begin{align}
 &\tilde{F}^{O(N)}_{1/N}|_{d=4-\eps}=\frac{1}{N}\left(\frac{7 \pi  \epsilon ^3}{288}+\frac{5 \pi  \epsilon ^4}{1728}-\frac{  (239+42 \pi ^2) \pi\epsilon ^5}{41472}+\mathcal{O}(\eps^{6})\right)\,, \notag\\
 &\tilde{F}^{O(N)}_{1/N}|_{d=6-\eps}=\frac{1}{N}\left(-\frac{7 \pi  \epsilon ^2}{120}+\frac{17 \pi  \epsilon ^3}{400}+\mathcal{O}(\eps^{4})\right)\,, \label{ONcritepsexp}\\
 & \tilde{F}^{O(N)}_{1/N}|_{d=8-\eps}=\frac{1}{N}\left(-\frac{215 \pi  \epsilon ^2}{1008}+\mathcal{O}(\eps^{3})\right)\,. \notag
\end{align}
One can check that the $d=4-\eps$ expansion in (\ref{ONcritepsexp}) exactly agrees with the result in  \cite{Fei:2015oha} and $d=6-\eps$ and $d=8-\eps$
expansions agree with formulas (\ref{FcubONres}) and (\ref{dF82res}).

\section*{Acknowledgments}

We are especially grateful to Igor Klebanov for suggesting this project and for his continued assistance and for careful reading of the manuscript. We are also grateful to Simone Giombi for many useful conversations, comments and suggestions. We thank Lin Fei for useful comments and input
to some calculations.  We also thank Aaron Levy and Silviu Pufu for useful discussions.
This work  was supported in part by the US NSF under Grant No.~PHY-1620059
 and a Myhrvold-Havranek Innovative Thinking Fellowship.

\appendix
\section{Large $N$ computation for the GN model}

In this appendix we present a direct computation  of the $1/N$ term in the sphere free energy of the Gross-Neveu model at general dimension. This computation 
assumes a special regularization of the Feynman integrals on a sphere. The computation of the 
integrals   is performed with the use of the Mellin-Barnes method \cite{Czakon:2005rk, Smirnov:2009up, Smirnov:2012gma}. Surprisingly it is possible to find an exact answer 
for arbitrary $d$. Unfortunately this answer is off by a factor of $1/3$ from the correct result. The origin of this discrepancy is unclear to us.

We start with the action describing critical fermion theory on a sphere in the large $N$ approach
\begin{equation}
S_{\textrm{GN}}= \int d^{d}x \sqrt{g } \Big(\bar{\psi}_{i}\slashed{\nabla}\psi^{i}+\frac{1}{\sqrt{N}}\sigma \bar{\psi}_{i} \psi^{i} \Big)\,,
\end{equation}
where $\psi^{i}$, $i=1,\ldots,N_{f}$ are Dirac fermions, $\sigma$ is the auxiliary field and $N =N_{f}{\rm Tr}{\bf 1}$.  So we have for the partition function 
 \begin{align}
Z_{\textrm{GN}} = Z_{f}^{N} \int D\sigma\frac{ \int D \psi  e^{- \int d^{d}x 
\sqrt{g}\frac{1}{\sqrt{N}}\sigma \bar{\psi}_{i} \psi^{i} } \exp(-S_{\textrm{free ferm}})}{\int D \psi \exp (-S_{\textrm{free ferm}}) } \,,
\end{align}
where $ Z_{f}^{N} =\int D \psi \exp (-S_{\textrm{free ferm}}) =\exp(-NF_{f})$ is  the partition function for the $N_{f}$ free Dirac fermions on the sphere and $F_{f}$ is given in (\ref{freeff}). Since the trace of odd number of gamma matrices is zero we have $\langle \bar{\psi}\psi(x_{1}) \dots  \bar{\psi}\psi(x_{2n+1})\rangle_{0} =0$. Therefore the $1/N$ expansion for the partition function takes the form
\begin{align}
&Z_{\textrm{GN}} = Z_{f}^{N}\int D\sigma \exp\Big(\frac{1}{2!N}\int d^{d}x_{1} d^{d}x_{2} \sqrt{g_{x_{1}}}  \sqrt{g_{x_{2}}} \sigma(x_{1})\sigma(x_{2}) \langle \bar{\psi}\psi(x_{1})\bar{\psi}\psi(x_{2})\rangle_{0} \notag\\
&~~~+\frac{1}{4! N^{2}} \int d^{d}x_{1} \sqrt{g_{x_{1}}} ...d^{d}x_{4} \sqrt{g_{x_{4}}} \sigma(x_{1})\dots \sigma(x_{4}) \langle \bar{\psi}\psi(x_{1})...\bar{\psi}\psi(x_{4})\rangle_{0}^{\textrm{conn}}+\dots \Big)\,. \label{GNpart1}
\end{align} 
The propagators in  flat space are related to those on the sphere  by the Weyl transformation, therefore
 we can compute correlation functions working in the flat space and then project the result on the sphere.
The propagators of the $\psi$  and $\sigma$ fields  in flat space are 
\begin{align}
\langle \psi^{i}(x) \bar{\psi}_{j}(0) \rangle_{0} &=- \delta^{i}_{j} \frac{\Gamma(\frac{d}{2})}{2\pi^{d/2}}\frac{\gamma^{\mu}x_{\mu}}{(x^{2})^{d/2}}=\delta^{i}_{j} C_{\psi}\frac{\slashed{x}}{(x^{2})^{d/2}}\,, \notag\\
\left<\sigma(x)\sigma(0) \right>_{0}&=-\frac{2^d \sin \left(\frac{\pi  d}{2}\right) \Gamma \left(\frac{d-1}{2}\right)}{\pi ^{3/2} \Gamma \left(\frac{d}{2}-1\right)} \frac{1}{(x^{2})^{1-\delta}}=\frac{C_{\sigma}}{(x^{2})^{1-\delta}}\,,
\label{sigma-eff-GN}
\end{align}
where  we have introduced a regulator $\delta$.  Next we define $ \Gamma_{2}(x_{1},x_{2})$ and $\Gamma_{4}(x_{1},...,x_{4})$ as  
\begin{align}
& \Gamma_{2}(x_{1},x_{2})\equiv \langle \bar{\psi}\psi(x_{1})\bar{\psi}\psi(x_{2})\rangle_{0} =  N_{f} C_{\psi}^{2} (-1) \frac{\Tr (\slashed{x}_{12}\slashed{x}_{21})}{(x_{12}^{2})^{d}}\,, \notag\\
&\Gamma_{4}(x_{1},...,x_{4})\equiv \langle \bar{\psi}\psi(x_{1})\dots \bar{\psi}\psi(x_{4})\rangle_{0}^{\textrm{conn}}   = 6N_{f}  C_{\psi}^{4}(-1) \frac{\Tr (\slashed{x}_{12}\slashed{x}_{23}\slashed{x}_{34}\slashed{x}_{41})}{(x_{12}^{2}x_{23}^{2}x_{34}^{2}x_{41}^{2})^{d/2}}\,, \label{Gammas1}
\end{align}
where $x_{ij}\equiv x_{i}-x_{j}$ and in the second line we assumed that we can symmetrize over the points $x_{1},\dots, x_{4}$ under the integral. Taking traces over the gamma matrices in (\ref{Gammas1})  we find 
\begin{align}
& \Gamma_{2}(x_{1},x_{2}) =  \frac{NC_{\psi}^{2} }{(x_{12}^{2})^{d-1}}\,, \quad \Gamma_{4}(x_{1},...,x_{4})=- 3NC_{\psi}^{4}\frac{(2x_{12}^{2}x_{34}^{2}-x_{13}^{2}x_{24}^{2})}{(x_{12}^{2}x_{23}^{2}x_{34}^{2}x_{41}^{2})^{d/2}}    \,,
\end{align}
where in the expression for $\Gamma_{4}$ we again used symmetrization over $x_{1},\dots, x_{4}$. In order  to write these correlation functions on the sphere one simply changes $x_{ij}\to s(x_{i},x_{j})$, where $s(x,y)$ is the chordal distance on the sphere. Now the partition function (\ref{GNpart1}) can be written in the form
\begin{align}
&Z_{\textrm{GN}} =Z_{f}^{N} \delta Z_{\Delta =d-1} \notag\\
&~~~~~~\times \Big\langle  \exp\Big(\frac{1}{4! N^{2} } \int d^{d}x_{1}\sqrt{g_{x_{1}}}...d^{d}x_{4}\sqrt{g_{x_{4}}}\,\sigma(x_{1})\dots\sigma(x_{4})\Gamma_{4}(x_{1},...,x_{4})+\dots\Big) \Big\rangle_{\sigma }\,, \label{GNpart2}
\end{align}
where $ \delta Z_{\Delta=d-1} =\exp(-\delta F_{\Delta=d-1}) = \int D\sigma \exp\big(\frac{1}{2N}\int d^{d}x_{1} d^{d}x_{2} \sigma(x_{1})\sigma(x_{2})\Gamma_{2}(x_{12}) \big)$ is the partition function on the sphere for the induced $\sigma$ field  and  $\delta F_{\Delta= d-1}$ is given by (\ref{Finduce}) with $\Delta =d-1$. The average in (\ref{GNpart2}) is taken over the $\sigma$ field 
with propagator (\ref{sigma-eff-GN})\footnote{The propagator of the $\sigma$ field on the sphere has the form $\langle \sigma(x)\sigma(y)\rangle= C_{\sigma}/(s(x,y))^{1-\delta}$.}. Finally for the sphere free energy  we obtain 
\begin{align}
&F_{\textrm{GN}} =NF_{f} +\delta F_{\Delta=d-1}\notag\\
&~~~~~~~~~ -\frac{1}{4!N^{2}}\int d^{d}x_{1}\sqrt{g_{x_{1}}}...d^{d}x_{4}\sqrt{g_{x_{4}}}\langle \sigma(x_{1})\dots \sigma(x_{4})\rangle_{\sigma} \Gamma_{4}(x_{1},...,x_{4})+\mathcal{O}(1/N^{2})\, . \label{FGNcrit1}
\end{align}
To compute the integral in (\ref{FGNcrit1}) we use the Mellin-Barnes approach, which is described in \cite{Fei:2015oha, Giombi:2015haa}. Surprisingly it is possible to compute this integral for arbitrary $d$ and the result does not have terms divergent  in  $\delta$ 
\begin{align}
\frac{1}{4! N^{2}}\int d^{d}x_{1}\sqrt{g_{x_{1}}}\dots d^{d}x_{4} \sqrt{g_{x_{4}}}\langle \sigma(x_{1})\dots \sigma(x_{4})\rangle_{\sigma} \Gamma_{4}(x_{1},\dots, x_{4})=-\frac{1}{N}\frac{3}{d} \gamma_{\psi, 1}^{\textrm{GN}}\,, \label{largenint}
\end{align}
where $\gamma_{\psi, 1}^{\textrm{GN}}$ is the $1/N$ correction to the anomalous dimension of $\psi$ field
\begin{align}
\gamma_{\psi, 1}^{\textrm{GN}} = \frac{(d-2)^2 \Gamma (d-1)}{4 \Gamma \left(2-\frac{d}{2}\right) \Gamma \left(\frac{d}{2}+1\right) \Gamma \left(\frac{d}{2}\right)^2}\,.
\end{align}
Nevertheless the result (\ref{largenint}) is off by a factor $1/3$ from the correct answer
\begin{align}
F_{\textrm{GN}}= NF_{f} + \delta F_{\Delta =d-1}
+\frac{1}{N}\frac{1}{d} \gamma_{\psi,1}^{\textrm{GN}}+\mathcal{O}(1/N^{2})\,.
\end{align}

The results for the Nambu-Jona-Lasinio models with $U(1)$ ans $SU(2)$ chiral symmetries in (\ref{Fres1}) are obtained using the same method  as described above for the Gross-Neveu model. The resulting Feynman  integrals are similar to the GN integrals  and again we need to multiply the final answer by the factor $1/3$.

\section{Large $N$ computation for the $O(N)$ model}

In this appendix we present a direct  calculation of the $1/N$ term in the sphere free energy of the $O(N)$ scalar model. The method of the computation is the same as in the Gross-Neveu model, discussed in previous section. It is also possible to find an exact answer 
for arbitrary $d$. Unfortunately this answer is off by some rational factors  from the correct result as in the Gross-Neveu case.

\begin{figure}[h!]
                \centering
                \includegraphics[width=12cm]{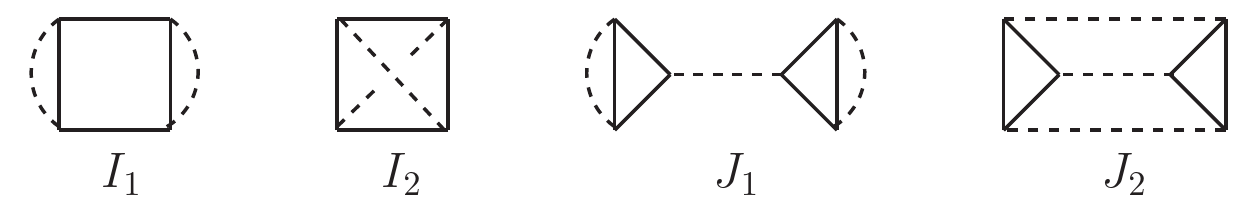}
                \caption{Diagrams contributing to the sphere free energy at  the $1/N$ order of the critical $O(N)$ scalar model. The $\langle \sigma \sigma\rangle$ and $\langle \phi \phi\rangle$ propagators are represented by the dashed and solid lines.}
                \label{diags1}
\end{figure}

We start with the action describing critical boson theory on the sphere in the large $N$ approach 
\begin{align}
S_{O(N) } = \int d^{d}x \sqrt{g} \bigg(\frac{1}{2}(\partial_{\mu} \phi^{i})^{2}+\frac{d-2}{8(d-1)} \mathcal{R}\phi^{i}\phi^{i}+\frac{1}{2\sqrt{N}}\sigma \phi^{i}\phi^{i}\bigg)\,,
\end{align}
where $\phi^{i}$, $i=1,\dots, N$ are scalar fields, $\sigma$ is the auxiliary scalar field and $\mathcal{R}$ is the scalar curvature.  For the partition function we get 
\begin{align}
Z _{O(N) }= Z_{s}^{N}\int D\sigma \frac{\int D\phi e^{-\int d^{d}x \sqrt{g} \frac{1}{2\sqrt{N}}\sigma \phi^{i}\phi^{i}}\exp(-S_{\textrm{free bos}})}{\int D\phi \exp(-S_{\textrm{free bos}})}\,, \label{ONpartF1}
\end{align}
where $ Z_{s}^{N} =\int D \phi \exp (-S_{\textrm{free bos}}) =\exp(-NF_{s})$ is  the partition function for the $N$ conformally coupled free bosons on the sphere and $F_{s}$ is given in (\ref{frees}). The propagators of the $\phi$ and $\sigma$ fields on the sphere are
\begin{align}
&\langle \phi^{i}(x)\phi^{j}(y)\rangle_{0} =
\delta^{ij} \frac{\Gamma(\frac{d}{2}-1)}{4\pi^{d/2}} \frac{1}{s(x,y)^{d-2}}=  \delta^{ij} \frac{C_{\phi}}{ s(x,y)^{d-2}}\,, \notag\\
&\langle \sigma(x) \sigma(y) \rangle =\frac{2^{d+2}\Gamma(\frac{d-1}{2})\sin(\frac{\pi d}{2})}{\pi^{\frac{3}{2}}\Gamma(\frac{d}{2}-2)} \frac{1}{s(x,y)^{2(2-\delta)}} = \frac{C_{\sigma}}{s(x,y)^{2(2-\delta)}}\,, \label{ONcritprops}
\end{align}
where we have introduced a regulator $\delta$. Expanding  the interaction term in (\ref{ONpartF1}) and 
defining correlation functions
\begin{align}
\Gamma_{2}(x_{1},x_{2})=&\langle \phi^{2}(x_{1})\phi^{2}(x_{2})\rangle_{0}=2NC_{\phi}^{2}(s(x_{1},x_{2}))^{-2(d-2)}\,, \notag\\
\Gamma_{3}(x_{1},x_{2},x_{3})=&\langle \phi^{2}(x_{1})\phi^{2}(x_{2})\phi^{2}(x_{3})\rangle^{\textrm{conn}}_{0} = 8NC_{\phi}^{3} (s(x_{12})s(x_{13})s(x_{23}))^{-(d-2)}\,, \\
\Gamma_{4}(x_{1},..., x_{4})= &\langle \phi^{2}(x_{1})\dots\phi^{2}(x_{4})\rangle^{\textrm{conn}}_{0} = 48NC_{\phi}^{4}(s(x_{12})s(x_{23})s(x_{34})s(x_{41}))^{-(d-2)}\,. \notag
\end{align}
where $\phi^{2}\equiv \phi^{i}\phi^{i}$ and $s(x_{ij})\equiv s(x_{i},x_{j})$, we obtain
\begin{align}
&Z _{O(N) }= Z_{s}^{N} \int D\sigma \exp\Big(\frac{1}{8N}\int d^{d}x_{1}d^{d}x_{2}\sqrt{g_{x_{1}}}\sqrt{g_{x_{2}}} \;\sigma(x_{1})\sigma(x_{2})\Gamma_{2}(x_{1},x_{2})\notag\\
&-\frac{1}{48N^{3/2}}\int d^{d}x_{1}\sqrt{g_{x_{1}}}...d^{d}x_{3}\sqrt{g_{x_{3}}} \sigma(x_{1})\sigma(x_{2})\sigma(x_{3})\Gamma_{3}(x_{1},x_{2},x_{3})\notag\\
&~~~~+\frac{1}{3\cdot 2^{7}N^{2}}\int d^{d}x_{1}\sqrt{g_{x_{1}}}...d^{d}x_{4}\sqrt{g_{x_{4}}} \sigma(x_{1})\sigma(x_{2})\sigma(x_{3})\sigma(x_{4})\Gamma_{4}(x_{1},...,x_{4})+\dots\Big) \,.
\end{align}
Finally taking average over the $\sigma$ field with the propagator (\ref{ONcritprops}) we find for the free energy 
\begin{align}
F _{O(N) }&=NF_{s}+\delta F_{\Delta=d-2} -\frac{1}{N}\Big(\frac{1}{4}I_{1}+\frac{1}{8}I_{2}+\frac{1}{8}J_{1}+\frac{1}{12}J_{2}\Big)+\mathcal{O}(1/N^{2})\,, \label{ONdeltaFcomb}
\end{align}
where $\delta F_{\Delta =d-2}$ is the sphere free energy of the induced $\sigma$ field and is given by (\ref{Finduce}) with $\Delta =d-2$, and  the integrals $I_{1}, I_{2}$ and $J_{1}, J_{2}$ read 
 \begin{align}
I_{1}&=\int d^{d}x_{1}\sqrt{g_{x_{1}}}\dots d^{d}x_{4} \sqrt{g_{x_{4}}}\frac{C_{\sigma}^{2}C_{\phi}^{4}}{(s(x_{12})s(x_{34}))^{2(2-\delta)}(s(x_{12})s(x_{23})s(x_{34})s(x_{41}))^{d-2}}\,,\notag\\
I_{2}&=\int d^{d}x_{1}\sqrt{g_{x_{1}}}\dots d^{d}x_{4} \sqrt{g_{x_{4}}}  \frac{C_{\sigma}^{2}C_{\phi}^{4}}{(s(x_{13})s(x_{24}))^{2(2-\delta)}(s(x_{12})s(x_{23})s(x_{34})s(x_{41}))^{d-2}}\, \label{IintON}
\end{align}
and also 
\begin{align}
J_{1} &=  \int  d^{d}x_{i}\sqrt{g_{x_{i}}} \frac{C_{\sigma}^{3}C_{\phi}^{6}}{(s(x_{12})s(x_{45}))^{2(d/2+1-\delta)}
s(x_{36})^{2(2-\delta)}
(s(x_{23})s(x_{31})s(x_{56})s(x_{64}))^{d-2}}\,, \notag\\
J_{2}&= \int  d^{d}x_{i}\sqrt{g_{x_{i}}}  \frac{C_{\sigma}^{3}C_{\phi}^{6}}{(s(x_{14})s(x_{25})s(x_{36}))^{2(2-\delta)}(s(x_{12})s(x_{23})s(x_{31})
s(x_{45})s(x_{56})s(x_{64}))^{d-2}}\,, \label{JintON}
\end{align}
where the diagrams for these integrals are represented in figure \ref{diags1}\,.
To compute the integrals in (\ref{IintON}) and (\ref{JintON}) we use the Mellin-Barnes approach.  As in the Gross-Neveu case it is also possible to compute these integrals for arbitrary $d$, and the result does not have terms divergent in  $\delta$. We obtain 
\begin{align}
&I_{1}=\Big(\frac{18}{d-1}-\frac{12}{d}-\frac{6}{d-2}\Big)\gamma_{\phi, 1}^{\textrm{O(N)}}\,, \quad I_{2}=\Big(\frac{12}{d-1}-\frac{12}{d-2}\Big)\gamma_{\phi, 1}^{\textrm{O(N)}}\,,\notag\\
&J_{1} =0\,,\quad J_{2}=\Big(\frac{20}{d-2}-\frac{40}{d-1}\Big)\gamma_{\phi,1}^{\textrm{O(N)}}\,, \label{I1I2J1J2}
\end{align}
where $\gamma_{\phi, 1}^{\textrm{O(N)}}$ is the $1/N$ correction to the anomalous dimension of the field $\phi$
\begin{align}
\gamma_{\phi, 1}^{\textrm{O(N)}} = \frac{2 \sin (\frac{\pi  d}{2}) \Gamma (d-2)}{\pi  \Gamma (\frac{d}{2}-2) \Gamma (\frac{d}{2}+1)}\,.
\end{align}
To get the correct  result for the sphere free energy we again, as in the Gross-Neveu case, we need to multiply the integrals  in (\ref{ONdeltaFcomb}) by some rational numbers  
\begin{align}
F _{O(N) }&=NF_{s}+\delta F_{\Delta =d-2} -\frac{1}{N}\left(\frac{1}{3}\Big(\frac{1}{4}I_{1}+\frac{1}{8}I_{2}\Big)+\frac{2}{5}\cdot\frac{1}{12}J_{2}\right)+\mathcal{O}(1/N^{2})\,.
\end{align}
Then  using (\ref{I1I2J1J2}) we find the final result
\begin{align}
F _{O(N)}&=NF_{s}+\delta F_{\Delta =d-2}+\frac{1}{N}\left(\frac{1}{d}+\frac{1}{3 (d-2)}-\frac{2}{3 (d-1)}\right)\gamma_{\phi,1}^{\textrm{O(N)}}+\mathcal{O}(1/N^{2})\,.
\end{align}

\bibliographystyle{ssg}
\bibliography{FoneoverN}

\end{document}